\documentclass[pra,aps,twocolumn,showpacs]{revtex4}

\usepackage{bm}
\usepackage{amsmath}
\usepackage{graphicx}
\usepackage{dcolumn}
\usepackage{bm}
\usepackage{psfrag}

\renewcommand{\Im}{\mathrm{Im}\,}
\newcommand{\grad}{\bm{\nabla}}

\newcommand{\vect}[1]{\mathbf{#1}}
\newcommand{\ten}[1]{\bm{#1}}

\newcommand{\tenszero}{\mbox{\textbf{\textit{0}}}}
\newcommand{\sprod}{\cdot}
\newcommand{\tprod}{}
\newcommand{\vprod}{\times}
\newcommand{\trace}{\operatorname{tr}}
\newcommand{\trans}{\mathsf{T}}
\newcommand{\dif}{\mathrm{d}}
\newcommand{\mi}{\mathrm{i}}
\newcommand{\me}{\mathrm{e}}

\begin{document}

\author{Stefan Scheel}
\email{s.scheel@imperial.ac.uk}
\author{Stefan Yoshi Buhmann}

\title{Casimir--Polder forces on moving atoms}

\affiliation{Quantum Optics and Laser Science, Blackett Laboratory,
Imperial College London, Prince Consort Road, London SW7 2AZ}

\begin{abstract}
Polarisable atoms and molecules experience the Casimir--Polder force
near magnetoelectric bodies, a force that is induced by quantum
fluctuations of the electromagnetic field and the matter. Atoms and
molecules in relative motion to a magnetoelectric surface experience
an additional, velocity-dependent force. We present a full
quantum-mechanical treatment of this force and identify a generalised
Doppler effect, the time delay between photon emission and
reabsorption, and the R\"{o}ntgen interaction as its three sources.
For ground-state atoms, the force is very small and always
decelerating, hence commonly known as quantum friction. For atom and
molecules in electronically excited states, on the contrary, both
decelerating and accelerating forces can occur depending on the
magnitude of the atomic transition frequency relative to the surface
plasmon frequency.

\end{abstract}
\date{\today}
\maketitle

%
\section{Introduction}
\label{Sec1}

The ground-state fluctuations of the electromagnetic field lead to
several inherently quantum effects such as the spontaneous decay of
excited atoms and molecules as well as dispersion forces
\cite{Milonni}. Forces between isolated atoms that are mediated by the
quantum vacuum are known as van~der~Waals forces \cite{vdW}, while
forces between macroscopic bodies are referred to as Casimir forces
\cite{Casimir48}. The third type of dispersion (in a sense an
interpolation between these two extreme cases) is the Casimir--Polder
(CP) force exerted on single atoms near macroscopic bodies
\cite{CasimirPolder}. 

For a two-level atom with transition frequency $\omega_A$ and electric
dipole moment $d$ located at a distance $z_A$ away from a perfectly
conducting plate, the short-distance (non-retarded)
\cite{CasimirPolder} and long-distance (retarded) \cite{LennardJones}
limits of the CP potential take the well-known forms 
\begin{equation}
U_\mathrm{nret}(z_A) =
-\frac{d^2}{48\pi\varepsilon_0z_A^3} \,,\quad
U_\mathrm{ret}(z_A) =
-\frac{cd^2}{16\pi^2\varepsilon_0\omega_A z_A^4}\,.
\end{equation}
These potentials, acting on atoms at rest, lead to conservative forces
perpendicular to the plate's surface. CP forces (as well as all other
dispersion forces) play important roles as limiting factors in efforts
to miniaturise atom-optical devices \cite{Fermani07}, and have been
measured at distances as small as $6\mu$m \cite{Harber05}.

Casimir--Polder forces are well understood far beyond the
aforementioned special case of a perfectly conducting plate, with
magnetoelectric materials of arbitrary shape \cite{Buhmann04} and
finite temperature being investigated theoretically
\cite{Antezza04,Buhmann08} as well as experimentally \cite{Obrecht07}.
While most theoretical investigations are based upon Lifshitz'
macroscopic treatment \cite{Lifshitz} or a linear-response description
\cite{McLachlan}, full quantum theories based upon
electromagnetic-field quantisation in magnetoelectrics have also been
developed \cite{Buhmann04}. In the latter approach, the
operator-valued Lorentz force
\begin{equation}
\hat{\mathbf{F}} = \int \dif^3r \left[ \hat{\rho}_A(\mathbf{r})
\hat{\mathbf{E}}(\mathbf{r}) +\hat{\mathbf{j}}_A(\mathbf{r}) \times
\hat{\mathbf{B}}(\mathbf{r}) \right] 
\end{equation}
on the atomic charge and current densities due to the body-assisted
electromagnetic fields is computed. In the long-wavelength
approximation, it leads to the well-known expression
\begin{equation}
\hat{\mathbf{F}} = \grad_A \left[
\hat{\mathbf{d}}\cdot\hat{\mathbf{E}}(\mathbf{r}_A) \right]
+\frac{\dif}{\dif t} \left[ 
\hat{\mathbf{d}}\times\hat{\mathbf{B}}(\mathbf{r}_A) \right]
\end{equation}
with $\hat{\mathbf{d}}$ denoting the atomic electric dipole moment
operator.

Intuitively, dispersion forces can be understood as dipole-dipole
forces generated by spontaneous polarisation due to the electric-field
fluctuations. Therefore, CP forces on atoms at rest act either towards
(attractively) or away (repulsively) from the macroscopic body. For
atoms in motion, retardation delays the dipole response, and a force
component emerges along the direction of motion. In most cases, this
force component acts against the motion and is thus the origin of
quantum friction.

Quantum friction forces have traditionally been studied within a
linear-response formalism \cite{Dedkov,Dorofeyev,Volokitin,Hu}.
Evaluating the correlated quantum fluctuations of moving atom and
dielectrics, the friction force on ground-state atoms can be obtained
in this way. However, the predicted forces are typically very small.
For the stationary case it is well known that CP forces can be
resonantly enhanced for excited atoms \cite{Buhmann04,Wylie}. For such
nonequilibrium situations, linear-response methods cannot be applied
and a more detailed investigation of the atom-field dynamics becomes
necessary.

In this article, we develop a full quantum theory of the
velocity-dependent CP force. In particular, we will show that for
atoms and molecules in electronically excited states, both
decelerating and accelerating forces can occur depending on relative
magnitude of the frequencies of the atomic transition and the surface
plasmon. The article is organised as follows: After briefly reviewing
the formalism of macroscopic quantum electrodynamics in
Sec.~\ref{Sec2}, we study the atom-field dynamics in Sec.~\ref{Sec3.1}
before investigating the resulting force in Sec.~\ref{Sec3.2} and
applying our results to the quantum friction scenario in
Sec.~\ref{Sec3.3}. We illustrate the theory with representative
examples in Sec.~\ref{Sec4}, followed by a summary in
Sec.~\ref{Sec:conclusions}.

%
\section{Basic formulas}
\label{Sec2}
Let us assume an arbitrary arrangement of dispersing and absorbing
magnetoelectric bodies, characterised by their complex-valued,
Kramers-Kronig consistent permittivity $\varepsilon(\vect{r},\omega)$
and permeability $\mu(\vect{r},\omega)$. The Hamiltonian of the
quantum electromagnetic field and the bodies can be given as
(for a recent review, see Ref.~\cite{Acta})
\begin{equation}
\label{qf1}
\hat{H}_F
 =\sum_{\lambda=e,m}
 \int\dif^3r\int_0^\infty\dif\omega\,\hbar\omega\,
 \hat{\vect{f}}_\lambda^\dagger(\vect{r},\omega)
 \sprod\hat{\vect{f}}_\lambda(\vect{r},\omega)
\end{equation}
where the fundamental variables
$\hat{\vect{f}}_\lambda^\dagger(\vect{r},\omega)$ and
$\hat{\vect{f}}_\lambda(\vect{r},\omega)$ are creation and
annihilation operators for the elementary electric ($\lambda=e$) and
magnetic ($\lambda=m$) excitations of the system; they obey the
bosonic commutation relations
\begin{gather}
\label{qf2}
\Bigl[\vect{\hat{f}}_\lambda(\vect{r},\omega),
 \vect{\hat{f}}_{\lambda'}^\dagger(\vect{r}',\omega')\Bigr]
 =\delta_{\lambda\lambda'}
 \bm{\delta}(\vect{r}-\vect{r}')\delta(\omega-\omega').
\end{gather}
The electric and magnetic fields can be expanded in terms of the
fundamental variables according to
\begin{eqnarray}
\label{qf3}
\hat{\vect{E}}(\vect{r})
&=&\int_0^{\infty}\dif\omega\,
 \underline{\hat{\vect{E}}}(\vect{r},\omega)+\operatorname{H.c.}
 \nonumber\\
&=&\sum_{\lambda={e},{m}}
 \int\dif^3r'\int_0^{\infty}\dif\omega\,
 \ten{G}_\lambda(\vect{r},\vect{r}',\omega)
 \sprod\hat{\vect{f}}_\lambda(\vect{r}',\omega)
 \nonumber\\
&&+\operatorname{H.c.},\\
\label{qf4}
\hat{\vect{B}}(\vect{r})
&=&\int_0^{\infty}\dif\omega\,
 \underline{\hat{\vect{B}}}(\vect{r},\omega)+\operatorname{H.c.}
 \nonumber\\
&=&\sum_{\lambda={e},{m}}
 \int\dif^3r'\int_0^{\infty}\frac{\dif\omega}{\mi\omega}\,
 \vect{\nabla}\vprod
 \ten{G}_\lambda(\vect{r},\vect{r}',\omega)
 \sprod\hat{\vect{f}}_\lambda(\vect{r}',\omega)
 \nonumber\\
&&+\operatorname{H.c.}
\end{eqnarray}
with coefficients
\begin{align}
\label{qf5}
&\ten{G}_e(\vect{r},\vect{r}',\omega)
 =\mi\,\frac{\omega^2}{c^2}
 \sqrt{\frac{\hbar}{\pi\varepsilon_0}\,
 \operatorname{Im}\varepsilon(\vect{r}',\omega)}\,
 \ten{G}(\vect{r},\vect{r}',\omega),\\
\label{qf6}
&\ten{G}_m(\vect{r},\vect{r}',\omega)
 =\mi\,\frac{\omega}{c}
 \sqrt{\frac{\hbar}{\pi\varepsilon_0}\,
 \frac{\operatorname{Im}\mu(\vect{r}',\omega)}
 {|\mu(\vect{r}',\omega)|^2}}
 \bigl[\vect{\nabla}'
 \!\!\times\!\ten{G}(\vect{r}',\vect{r},\omega)
 \bigr]^{\trans}.
\end{align}
Here, $\ten{G}$ is the classical Green tensor as uniquely defined by
the inhomogeneous Helmholtz equation
\begin{equation}
\label{qf7}
\left[\bm{\nabla}\times
\frac{1}{\mu(\vect{r},\omega)}\bm{\nabla}\times
 \,-\,\frac{\omega^2}{c^2}\,\varepsilon(\vect{r},\omega)\right]
 \ten{G}(\vect{r},\vect{r}',\omega)
 =\bm{\delta}(\vect{r}-\vect{r}')
\end{equation}
together with the boundary condition
\begin{equation}
\label{qf8}
\ten{G}(\vect{r},\vect{r}',\omega)\to \tenszero
\quad\mbox{for }|\vect{r}-\vect{r}'|\to\infty.
\end{equation}
The Green tensor is an analytic function in the upper half of the
complex frequency plane and it has the following useful properties:
\begin{gather}
\label{qf9}
\ten{G}(\vect{r},\vect{r}',-\omega^\ast) 
 =\ten{G}^\ast(\vect{r},\vect{r}',\omega),\\
\label{qf10}
\ten{G}(\vect{r}',\vect{r},\omega) 
 =\ten{G}^\trans(\vect{r},\vect{r}',\omega),\\
\label{qf11}
\sum_{\lambda={e},{m}}\int\dif^3 s\,
 \ten{G}_\lambda(\vect{r},\vect{s},\omega)\!\cdot\!
 \ten{G}^{\ast\trans}_\lambda(\vect{r}',\vect{s},\omega)\\
\qquad=\frac{\hbar\mu_0}{\pi}\,\omega^2\operatorname{Im}
 \ten{G}(\vect{r},\vect{r}',\omega).
\end{gather}

The Hamiltonian describing the internal dynamics of an atom with
eigenenergies $E_n$ and eigenstates $|n\rangle$ can be
given as
\begin{equation}
\label{qf12}
\hat{H}_A
 =\sum_k E_k\hat{A}_{kk}
\end{equation}
($\hat{A}_{kk}=|k\rangle\langle k|$: atomic flip operators).
Throughout this article, we will assume that the centre-of-mass motion
is sufficiently slow so that it separates from the internal dynamics
in the spirit of a Born--Oppenheimer approximation. The interaction of
the atom with the body-assisted electromagnetic field is then
adequately described by the atom-field coupling Hamiltonian for given
centre-of mass position $\vect{r}_A$ and velocity $\vect{v}_A$, which
in the multipolar coupling scheme and electric-dipole approximation
reads
\begin{align}
\label{qf13}
\hat{H}_{AF}
=&-\hat{\vect{d}}\sprod\hat{\vect{E}}(\vect{r}_A)
 -\hat{\vect{d}}\sprod\vect{v}_A\vprod\hat{\vect{B}}(\vect{r}_A)
 \nonumber\\
=&-\sum_{kl}\vect{d}_{kl}\sprod\hat{\vect{E}}(\vect{r}_A)
 \hat{A}_{kl}
 -\sum_{kl}\vect{d}_{kl}\sprod\vect{v}_A
 \vprod\hat{\vect{B}}(\vect{r}_A)\hat{A}_{kl}.\nonumber\\
\end{align}
The first term is the familiar electric-dipole interaction while the
second term is the R\"{o}ntgen interaction associated with the
centre-of-mass motion. Combining Eqs.~(\ref{qf1}), (\ref{qf12}), and
(\ref{qf13}), the total Hamiltonian of the atom--body--field system
reads
\begin{equation}
\label{qf14}
\hat{H}=\hat{H}_A+\hat{H}_F+\hat{H}_{AF}.
\end{equation}

Finally, the total Lorentz force on the atomic charge and current
distribution can in electric-dipole approximation be given as
\begin{align}
\label{qf15}
\hat{\vect{F}}=&\,\bm{\nabla}_A \bigl[
\hat{\vect{d}}\sprod\hat{\vect{E}}(\vect{r}_A)
+\hat{\vect{d}}\sprod\vect{v}_A\vprod
\hat{\vect{B}}(\vect{r}_A)\bigr] 
+\frac{\dif}{\dif t} \bigl[\hat{\vect{d}}\vprod
\hat{\vect{B}}(\vect{r}_A)\bigr]\nonumber\\
=&\,\bm{\nabla}_A\sum_{kl} \bigl[
\vect{d}_{kl}\sprod\hat{\vect{E}}(\vect{r}_A)\hat{A}_{kl}
+\vect{d}_{kl}\sprod\vect{v}_A\vprod
\hat{\vect{B}}(\vect{r}_A)\hat{A}_{kl}\bigr]\nonumber\\ 
&\,+\frac{\dif}{\dif t} \sum_{kl}\bigl[\vect{d}_{kl}\vprod
\hat{\vect{B}}(\vect{r}_A)\hat{A}_{kl}\bigr].
\end{align}
%
%
\section{Casimir--Polder force on a moving atom}
\label{Sec3}
The Casimir--Polder force on an atom is the quantum average of the
Lorentz force~(\ref{qf15}) with the body-assisted field being in its
ground state. To evaluate this expression, we first need to solve the
coupled atom--field dynamics.
%
%
\subsection{Atom--field dynamics}
\label{Sec3.1}
Using the Hamiltonian~(\ref{qf14}), the Heisenberg equations of motion
of the atomic and field operators are found to be
\begin{multline}
\label{qf16}
\dot{\hat{A}}_{mn}
 =\mi\omega_{mn} \hat{A}_{mn}
 +\frac{\mi}{\hbar}\sum_k
 \bigl(\vect{d}_{nk}\hat{A}_{mk}-\vect{d}_{km}\hat{A}_{kn}\bigr)
 \sprod\hat{\vect{E}} (\hat{\vect{r}}_{\!A})\\
+\frac{\mi}{\hbar}\sum_k
 \bigl(\vect{d}_{nk}\hat{A}_{mk}-\vect{d}_{km}\hat{A}_{kn}\bigr)
 \sprod\vect{v}_A
 \vprod\hat{\vect{B}} (\hat{\vect{r}}_{\!A})
\end{multline}
and
\begin{multline}
\label{qf17}
\dot{\hat{\vect{\!f}}}_\lambda(\vect{r},\omega)
=-\mi\omega\hat{\vect{f}}_\lambda(\vect{r},\omega)
 +\frac{\mi}{\hbar}\sum_{k,l}
 \ten{G}_\lambda^{\ast\trans}(\vect{r},\vect{r}_A,\omega)
 \sprod\vect{d}_{kl}\hat{A}_{kl}\\
+\frac{1}{\hbar\omega}\sum_{k,l}
 \bigl[\ten{G}_\lambda^{\ast\trans}(\vect{r},\vect{r}_A,\omega)
 \vprod\overleftarrow{\bm{\nabla}}'\bigr]\vprod\vect{v}_A
 \sprod\vect{d}_{kl}\hat{A}_{kl}
\end{multline}
(by convention, $\bm{\nabla}$ and $\bm{\nabla}'$ only act on the
first or second argument of the Green tensor, respectively).
The latter equation is formally solved by
\begin{equation}
\label{qf18}
\hat{\vect{f}}_\lambda(\vect{r},\omega,t)
=\hat{\vect{f}}_{\lambda,\mathrm{f}}(\vect{r},\omega,t)
 +\hat{\vect{f}}_{\lambda,\mathrm{s}}(\vect{r},\omega,t)
\end{equation}
where
\begin{align}
\label{qf19}
&\hat{\vect{f}}_{\lambda,\mathrm{f}}(\vect{r},\omega,t)
 =\me^{-\mi\omega t}
 \hat{\vect{f}}_\lambda(\vect{r},\omega),\\
\label{qf20}
&\hat{\vect{f}}_{\lambda,\mathrm{s}}(\vect{r},\omega,t)\nonumber\\
&=\frac{\mi}{\hbar}\sum_{k,l}\int_0^t\dif\tau\,
 \me^{-\mi\omega(t-\tau)}\ten{G}_\lambda^{\ast\trans}
 [\vect{r},\vect{r}_{\!A}(\tau),\omega]\sprod
 \vect{d}_{kl}\hat{A}_{kl}(\tau)\nonumber\\
&\quad-\frac{1}{\hbar\omega}\sum_{k,l}\int_0^t\dif\tau\,
 \me^{-\mi\omega(t-\tau)}\bigl\{\ten{G}_\lambda^{\ast\trans}
 [\vect{r},\vect{r}_{\!A}(\tau),\omega]\vprod
 \overleftarrow{\bm{\nabla}}'\bigr\}\nonumber\\
&\qquad\vprod\vect{v}_A\sprod
 \vect{d}_{kl}\hat{A}_{kl}(\tau)
\end{align}
determine the free and source parts of the electromagnetic field.

We assume that the atom moves with uniform, nonrelativistic speed
($v_A\ll c$) and we are seeking a solution to the system of
Eqs.~(\ref{qf16}) and (\ref{qf17}) within linear order of
$\vect{v}_A$. We may hence write
\begin{equation}
\label{qf21}
\vect{r}_{\!A}(\tau)=\vect{r}_{\!A}(t)-(t-\tau)\vect{v}_A;
\end{equation}
and after substituting Eqs.~(\ref{qf18})--(\ref{qf20}) into
Eq.~(\ref{qf3}), using the integral relation~(\ref{qf11}), and
applying a linear Taylor expansion in $\vect{v}_A$, the time-dependent
frequency components of the electric field are given by
\begin{equation}
\label{qf22}
\underline{\hat{\vect{E}}}(\vect{r},\omega,t)
=\underline{\hat{\vect{E}}}_\mathrm{f}(\vect{r},\omega,t)
 +\underline{\hat{\vect{E}}}_\mathrm{s}(\vect{r},\omega,t)
\end{equation}
with
\begin{align}
\label{qf23}
&\underline{\hat{\vect{E}}}_\mathrm{f}(\vect{r},\omega,t) 
 =\me^{-\mi\omega t}
 \underline{\hat{\vect{E}}}(\vect{r},\omega)\\
\label{qf25}
&\underline{\hat{\vect{E}}}_\mathrm{s}(\vect{r},\omega,t)
 \nonumber\\
&=\frac{\mi\mu_0}{\pi}\,\omega^2
 \int_0^t\dif\tau\,\me^{-\mi\omega(t-\tau)}
 \sum_{k,l}\mathrm{Im}\,
 \ten{G}(\vect{r},\vect{r}_A,\omega)\sprod
 \vect{d}_{kl}\hat{A}_{kl}(\tau)\nonumber\\
&\quad-\frac{\mi\mu_0}{\pi}\,\omega^2
 \int_0^t\dif\tau\,(t-\tau)\me^{-\mi\omega(t-\tau)}\nonumber\\
&\qquad\times\sum_{k,l}\mathrm{Im}\,
 \ten{G}(\vect{r},\vect{r}_A,\omega)\sprod
 \vect{d}_{kl}\bigl(\overleftarrow{\bm{\nabla}}'\sprod\vect{v}_A\bigr)
 \hat{A}_{kl}(\tau)\nonumber\\
&\quad-\frac{\mu_0}{\pi}\,\omega
 \int_0^t\dif\tau\,\me^{-\mi\omega(t-\tau)}
 \sum_{k,l}\mathrm{Im}\,
 \bigl[\ten{G}(\vect{r},\vect{r}_A,\omega)\vprod
 \overleftarrow{\bm{\nabla}}'\bigr]\nonumber\\
&\qquad\vprod\vect{v}_A
 \sprod\vect{d}_{kl}\hat{A}_{kl}(\tau)
\end{align}
[$\vect{r}_A=\vect{r}_A(t)$]. The magnetic field~(\ref{qf4}) only
enters the equations of motion in conjunction with a factor
$\vect{v}_A$, so we only require its zero-order expansion in the
velocity:
\begin{equation}
\label{qf26}
\underline{\hat{\vect{B}}}(\vect{r},\omega,t)
=\underline{\hat{\vect{B}}}_\mathrm{f}(\vect{r},\omega,t)
 +\underline{\hat{\vect{B}}}_\mathrm{s}(\vect{r},\omega,t)
\end{equation}
with
\begin{align}
\label{qf27}
\underline{\hat{\vect{B}}}_\mathrm{f}(\vect{r},\omega,t) 
 =&\,\me^{-\mi\omega t}
 \underline{\hat{\vect{B}}}(\vect{r},\omega)\\
\label{qf28}
\underline{\hat{\vect{B}}}_\mathrm{s}(\vect{r},\omega,t)
=&\,\frac{\mu_0}{\pi}\,\omega
 \int_0^t\dif\tau\,\me^{-\mi\omega(t-\tau)}\nonumber\\
&\times\sum_{k,l}\bm{\nabla}\vprod\mathrm{Im}\,
 \ten{G}(\vect{r},\vect{r}_A,\omega)\sprod
 \vect{d}_{kl}\hat{A}_{kl}(\tau).
\end{align}

We can next substitute our solutions~(\ref{qf23})--(\ref{qf28}) for
the time-independent electromagnetic fields into the equation of
motion~(\ref{qf16}) for the atomic flip operators. Noting that the
total field operators $\underline{\hat{\vect{E}}}(\vect{r},\omega,t)$
and $\underline{\hat{\vect{B}}}(\vect{r},\omega,t)$ commute with the
atomic flip operators at equal times, we arrange all products such
that creation operators
$\hat{\vect{f}}_\lambda^\dagger(\vect{r},\omega)$ are always at the
left and annihilation operators
$\hat{\vect{f}}_\lambda(\vect{r},\omega)$ are always at the right.
Assuming the field to be initially prepared in its vacuum state and
taking expectation values, all contributions from the source fields
vanish. For weak atom--field coupling, the time integrals can be
evaluated with the aid of the Markov approximation,
\begin{multline}
\label{qf29}
\int_0^t\dif\tau\,\me^{-\mi\omega(t-\tau)}
 \bigl\langle\hat{A}_{ij}(t)\hat{A}_{kl}(\tau)\bigr\rangle\\
\simeq\bigl\langle\hat{A}_{ij}(t)\hat{A}_{kl}(t)\bigr\rangle
\int_{-\infty}^t\dif\tau\,
 \me^{-\mi(\omega-\tilde{\omega}_{lk})(t-\tau)}\\
=\bigl\langle\hat{A}_{il}(t)\bigr\rangle\delta_{jk}
 \biggl[\pi\delta(\omega-\tilde{\omega}_{lk})
 -\mi\frac{\mathcal{P}}{\omega-\tilde{\omega}_{lk}}\biggr]
\end{multline}
($\mathcal{P}$: principal value); similarly we have
\begin{multline}
\label{qf30}
\int_0^t\dif\tau\,(t-\tau)\me^{-\mi\omega(t-\tau)}
 \bigl\langle\hat{A}_{ij}(t)\hat{A}_{kl}(\tau)\bigr\rangle\\
\simeq\bigl\langle\hat{A}_{il}(t)\bigr\rangle\delta_{jk}
 \frac{\dif}{\dif\omega}\biggl[
 \frac{\mathcal{P}}{\omega-\tilde{\omega}_{lk}}
 +\mi\pi\delta(\omega-\tilde{\omega}_{lk})\biggr],
\end{multline}
where the shifted atomic transition frequencies 
\begin{equation}
\label{qf31}
\tilde{\omega}_{mn}
=\omega_{mn}+\delta\omega_m-\delta\omega_n
\end{equation}
have yet to be determined. 

For a nondegenerate atom, the resulting equations of motion for the
off-diagonal atomic flip operators decouple from each other as well as
from the diagonal ones. In addition, we consider an atom whose
internal Hamiltonian~(\ref{qf12}) is time-reversal invariant, so
that we may assume real dipole-matrix elements. After a lengthy, but
straightforward calculation, we finally obtain the following equations
of motion for the internal atomic density matrix elements
$\sigma_{mn}=\langle\hat{A}_{nm}\rangle$:
\begin{gather}
\label{qf32}
\dot{p}_n
 =-\Gamma_np_n
 +\sum_k\Gamma_k^np_k,\\ 
\label{qf33}
\dot{\sigma}_{mn}
 =[-\mi\tilde{\omega}_{mn}
 -(\Gamma_m+\Gamma_n)/2]\sigma_{mn}
 \quad\mbox{for }m\neq n,
\end{gather}
where we have introduced the probabilities $p_n=\sigma_{nn}$. The
equations of motion for a moving atom have exactly the same form as
for an atom at rest: The population of the diagonal density matrix
elements is governed by spontaneous decay, while the off-diagonal ones
undergo damped oscillations. However, the respective transition rates 
\begin{equation}
\label{qf34}
\Gamma_n = \sum_k \Gamma_n^k
\end{equation}
and
frequency shifts
\begin{equation}
\label{qf35}
\delta\omega_n =\sum_k \delta\omega_n^k
\end{equation}
are affected by the atomic motion:
\begin{gather} 
\label{qf36}
\delta\omega_n^k=\delta\omega_n^k(\vect{r}_A)
 +\delta\omega_n^k(\vect{r}_A,\vect{v}_A),\\
\label{qf36b}
\delta\omega_n^k(\vect{r}_A)
=\frac{\mu_0}{\pi\hbar}\,
 \mathcal{P}\!\!\int_0^\infty\!\!\!\dif\omega\,
 \frac{\omega^2\vect{d}_{nk}\sprod\mathrm{Im}\,
 \ten{G}^{(1)}(\vect{r}_A,\vect{r}_A,\omega)\sprod
 \vect{d}_{kn}}{\tilde{\omega}_{nk}-\omega}\,,\\
\label{qf36c}
\delta\omega_n^k(\vect{r}_A,\vect{v}_A)=\frac{\mu_0}{2\hbar}\,
 \Theta(\tilde{\omega}_{nk})(\vect{v}_A\sprod\bm{\nabla}_A)
 \hspace{12ex}\nonumber\\
\hspace{8ex}\times\bigl[\omega^2\vect{d}_{nk}\sprod
 \mathrm{Im}\,\ten{G}(\vect{r}_A,\vect{r}_A,\omega)
 \sprod\vect{d}_{kn}\bigr]'_{\omega=\tilde{\omega}_{nk}},\\
\label{qf37}
\Gamma_n^k=\Gamma_n^k(\vect{r}_A)
 +\Gamma_n^k(\vect{r}_A,\vect{v}_A),\\
\label{qf37b}
\Gamma_n^k(\vect{r}_A)
=\frac{2\mu_0}{\hbar}\,\Theta(\tilde{\omega}_{nk})
 \tilde{\omega}_{nk}^2\vect{d}_{nk}\sprod
 \mathrm{Im}\,\ten{G}(\vect{r}_A,\vect{r}_A,
 \tilde{\omega}_{nk})\sprod\vect{d}_{kn},\\
\label{qf37c}
\Gamma_n^k(\vect{r}_A,\vect{v}_A)
=-\frac{\mu_0}{\pi\hbar}\,(\vect{v}_A\sprod\bm{\nabla}_A)
 \mathcal{P}\int_0^\infty\dif\omega\,\hspace{6ex}\nonumber\\
\hspace{2ex}\times\frac{\bigl[\omega^2\vect{d}_{nk}\sprod\mathrm{Im}\,
 \ten{G}^{(1)}(\vect{r}_A,\vect{r}_A,\omega)\sprod
 \vect{d}_{kn}\bigr]'}{\tilde{\omega}_{nk}-\omega}\,.
\end{gather}
(the primes indicate derivatives with respect to $\omega$). Here, we
have decomposed the Green tensor into its bulk (free-space) and
scattering parts according to
\begin{equation}
\label{qf39}
\ten{G}(\vect{r},\vect{r}',\omega)
=\ten{G}^{(0)}(\vect{r},\vect{r}',\omega)
+\ten{G}^{(1)}(\vect{r},\vect{r}',\omega)
\end{equation}
and have discarded the Lamb-shift contribution due to $\ten{G}^{(0)}$
from the frequency shift $\delta\omega_n^k(\vect{r}_A)$ (as the
free-space Lamb shift is assumed to be already included in the bare
transition frequencies $\omega_{mn}$). We have further exploited the
symmetry~(\ref{qf10}) of the Green tensor which implies
$\bm{\nabla}'\mapsto\frac{1}{2}\bm{\nabla}_A$, showing that the
translationally invariant bulk Green tensor does not contribute to the
velocity-dependent shifts and rates. 

It is worth noting that for real dipole matrix elements, the
contributions from the R\"{o}ntgen interaction exactly cancel. As a
result, the velocity dependence of these quantities is entirely due to
the fact that the moving atom emits and receives the electromagnetic
field at different positions. The velocity-dependent contributions are
proportional to the total derivative of the scattering Green tensor
along the direction of motion. As a consequence, the decay rates and
frequency shifts are unaffected by uniform motion in a direction along
which the environment is translationally invariant (e.g. motion
parallel to a plate or a cylinder). It is worth pointing out that
such a vanishing of velocity-dependent frequency shifts does not
necessarily imply that the velocity-dependent part of the CP force
must also be zero. One should bear in mind that all of the above has
only been shown within linear order in the velocity.
%
%
\subsection{Casimir--Polder force}
\label{Sec3.2}
Having solved the coupled atom--field dynamics, we can now evaluate
the quantum average of the Lorentz force~(\ref{qf15}). We restrict
our attention to the pure dispersion force by again assuming the field
to be initially prepared in its ground state. The atom may initially
be in an arbitrary incoherent superposition of internal energy
eigenstates. For an atom at rest, it has been found that the third
term in Eq.~(\ref{qf15}), which involves a total time derivative,
does not contribute to the force on atoms in incoherent internal
states. We have explicitly checked that the same is true here, so
that we only need to consider the force
\begin{align}
\label{qf40}
\vect{F}=&\,\bm{\nabla}_A \bigl\langle
\hat{\vect{d}}\sprod\hat{\vect{E}}(\vect{r}_A)
+\hat{\vect{d}}\sprod\vect{v}_A\vprod
\hat{\vect{B}}(\vect{r}_A)\bigr\rangle\nonumber\\
=&\,\bm{\nabla}_A\sum_{kl} \bigl[
\vect{d}_{kl}\sprod\bigl\langle\hat{\vect{E}}(\vect{r}_A)\hat{A}_{kl}
\bigr\rangle 
+\vect{d}_{kl}\sprod\vect{v}_A\vprod
\bigl\langle\hat{\vect{B}}(\vect{r}_A)\hat{A}_{kl}\bigr\rangle\bigr].
\end{align}

We begin by substituting the time-dependent electromagnetic
fields~(\ref{qf22})--(\ref{qf28}) where again we retain only terms up
to linear order in $\vect{v}_A$ and we arrange all products
such that the contributions from the free fields vanish. The source
fields give rise to intra-atomic correlation functions. By virtue of
the quantum regression theorem, Eq.~(\ref{qf33}) implies that the
relevant correlation functions are of the form
\begin{equation}
\label{qf41}
\bigl\langle\hat{A}_{nk}(t)\hat{A}_{ln}(\tau)\bigr\rangle
 =\delta_{kl}\me^{\mi\Omega_{nk}(t-\tau)}\sigma_{nn}(\tau) 
\end{equation}
with 
\begin{equation}
\label{qf42}
\Omega_{nk}=\tilde{\omega}_{nk}+\mi(\Gamma_n+\Gamma_k)/2.
\end{equation}
We evaluate time integrals in the spirit of the Markov approximation:
\begin{multline}
\label{qf43}
\int_0^t\dif\tau\,\me^{-\mi\omega(t-\tau)}
 \bigl\langle\hat{A}_{nk}(t)\hat{A}_{ln}(\tau)\bigr\rangle\\
\simeq\sigma_{nn}(t)\delta_{kl} 
\int_{-\infty}^t\dif\tau\,
 \me^{-\mi(\omega-\Omega_{nk})(t-\tau)}
=-\frac{\sigma_{nn}(t)\mi\delta_{kl}}{\omega-\Omega_{nk}}\;,
\end{multline}
and similarly
\begin{multline}
\label{qf44}
\int_0^t\dif\tau\,(t-\tau)\me^{-\mi\omega(t-\tau)}
 \bigl\langle\hat{A}_{nk}(t)\hat{A}_{ln}(\tau)\bigr\rangle\\
\simeq-\frac{\sigma_{nn}(t)\delta_{kl}}{(\omega-\Omega_{nk})^2}\,.
\end{multline}
Again assuming real dipole matrix elements, the resulting expression
for the CP force can be written in the form
\begin{equation}
\label{qf45}
\vect{F}(t)=\sum_np_n(t)\vect{F}_n
\end{equation}
with
\begin{widetext}
\begin{multline}
\label{qf46}
\vect{F}_n 
=\frac{\mu_0}{\pi}\sum_k\int_0^\infty\!\!\!\dif\omega\,\omega^2\,
 \frac{\bm{\nabla}\tprod\vect{d}_{nk}\sprod
 \mathrm{Im}\,\ten{G}^{(1)}(\vect{r}_A,\vect{r}_A,\omega)
 \sprod\vect{d}_{kn}}{\omega-\Omega_{nk}}
+\frac{\mi\mu_0}{\pi}\sum_k\int_0^\infty\!\!\!\dif\omega\,\omega^2\,
 \frac{\bm{\nabla}(\vect{v}_A\sprod\bm{\nabla}')
 \tprod\vect{d}_{nk}\sprod
 \mathrm{Im}\,\ten{G}^{(1)}(\vect{r}_A,\vect{r}_A,\omega)
 \sprod\vect{d}_{kn}}
 {(\omega-\Omega_{nk})^2}\\
+\frac{\mi\mu_0}{\pi}\sum_k\int_0^\infty\!\!\!\dif\omega\,\omega\,
 \frac{(\bm{\nabla}'-\bm{\nabla})\tprod\vect{d}_{nk}\sprod
 \vect{v}_A\vprod\bigl[\bm{\nabla}\vprod
 \mathrm{Im}\,\ten{G}^{(1)}(\vect{r}_A,\vect{r}_A,\omega)\bigr]
 \sprod\vect{d}_{kn}}{\omega-\Omega_{nk}}
 +\mathrm{C.c.}
\end{multline}
\end{widetext}
Note that the two contributions from the R\"{o}ntgen interaction have
been collected in a single term as given on the second line of the
above equation by making use of the symmetry~(\ref{qf10}) of the Green
tensor. In addition, the (vanishing) contributions from the free-space
Green tensor have been discarded.

Next, let us separate the forces $\vect{F}_n$ into their position-
and velocity- dependent parts. The shifted and broadened atomic
transition frequencies $\Omega_{nk}$ are velocity-dependent, so
that the first term in Eq.~(\ref{qf46}) also contributes to the
velocity-dependent part of the force. Again retaining only terms up
to linear order in the velocity, we find
\begin{equation}
\label{qf47}
\vect{F}_n 
=\vect{F}_n(\vect{r}_A)+\vect{F}_n(\vect{r}_A,\vect{v}_A),
\end{equation}
with
\begin{multline}
\label{qf48}
\vect{F}_n(\vect{r}_A)
=\frac{\mu_0}{2\pi}\sum_k\int_0^\infty\!\!\!\dif\omega\,\omega^2\\
\times\frac{\bm{\nabla}_A\tprod\vect{d}_{mk}\sprod
 \mathrm{Im}\,\ten{G}^{(1)}(\vect{r}_A,\vect{r}_A,\omega)
 \sprod\vect{d}_{kn}}{\omega-\Omega_{nk}}
 +\mathrm{C.c.}
\end{multline}
and
\begin{widetext}
\begin{multline}
\label{qf49}
\vect{F}_n(\vect{r}_A,\vect{v}_A)
=\frac{\mu_0}{\pi}\sum_k\int_0^\infty\dif\omega\,\omega^2\,
 \frac{\Omega_{nk}(\vect{v}_A)\bm{\nabla}\tprod\vect{d}_{nk}\sprod
 \mathrm{Im}\,\ten{G}^{(1)}(\vect{r}_A,\vect{r}_A,\omega)
 \sprod\vect{d}_{kn}}{(\omega-\Omega_{nk})^2}\\
+\frac{\mi\mu_0}{\pi}\sum_k\int_0^\infty\dif\omega\,\omega^2\,
 \frac{\bm{\nabla}(\vect{v}_A\sprod\bm{\nabla}')
 \tprod\vect{d}_{nk}\sprod
 \mathrm{Im}\,\ten{G}^{(1)}(\vect{r}_A,\vect{r}_A,\omega)
 \sprod\vect{d}_{kn}}
 {(\omega-\Omega_{nk})^2}\\
+\frac{\mi\mu_0}{\pi}\sum_k\int_0^\infty\dif\omega\,\omega\,
 \frac{(\bm{\nabla}'-\bm{\nabla})\tprod\vect{d}_{nk}\sprod
 \vect{v}_A\vprod\bigl[\bm{\nabla}\vprod
 \mathrm{Im}\,\ten{G}^{(1)}(\vect{r}_A,\vect{r}_A,\omega)\bigr]
 \sprod\vect{d}_{kn}}{\omega-\Omega_{nk}}
 +\mathrm{C.c.}
\end{multline}
\end{widetext}
where $\Omega_{nk}\equiv\Omega_{nk}(\vect{r}_A)$ and
$\Omega_{nk}(\vect{v}_A)\equiv\Omega_{nk}(\vect{r}_A,\vect{v}_A)$.
The velocity-independent force~(\ref{qf48}) is just the well-known CP
force on an atom at rest. We will in the following restrict our
attention to the velocity-dependent force~(\ref{qf48}), which
consists of three terms: The first, generalised Doppler term is due
to the velocity-dependence of the atomic transition frequencies; the
second, delay term is associated with the time interval between
emission and reabsorption of the electromagnetic field; and the
third, R\"{o}ntgen term is due to the coupling of the current density
associated with the moving atom to the magnetic field.

In close analogy to the case of an atom at rest, the force can be
separated into its resonant and nonresonant parts using
contour-integral techniques. Writing
$\mathrm{Im}\ten{G}=(\ten{G}-\ten{G}^\ast)/(2\mi)$, using the
property~(\ref{qf9}) of the Green tensor, and employing Cauchy's
theorem to transform integrals along the real axis to integrals along
the positive imaginary axis plus contributions from the poles, one
can show that
\begin{align}
\label{qf50}
&\int_0^\infty\dif\omega\,\frac{\omega\Im
 \ten{G}^{(1)}(\vect{r},\vect{r}',\omega)}
 {\omega-\Omega_{nk}}
=\int_0^\infty\dif\xi\,\frac{\xi^2
 \ten{G}^{(1)}(\vect{r},\vect{r}',\mi\xi)}
 {\xi^2+\Omega_{nk}^2}\nonumber\\
&\qquad+\pi\Omega_{nk}\ten{G}^{(1)}(\vect{r},\vect{r}',\Omega_{nk})
 \Theta(\tilde{\omega}_{nk}),
\end{align}
\begin{align}
\label{qf51}
&\int_0^\infty\dif\omega\,\frac{\omega^2\Im
 \ten{G}^{(1)}(\vect{r},\vect{r}',\omega)}
 {(\omega-\Omega_{nk})^2}\nonumber\\
&\quad=-\int_0^\infty\dif\xi\,\frac{\xi^2(\Omega_{nk}^2-\xi^2)
 \ten{G}^{(1)}(\vect{r},\vect{r}',\mi\xi)}
 {(\xi^2+\Omega_{nk}^2)^2}\nonumber\\
&\qquad+\pi\bigl[\omega^2\ten{G}^{(1)}(\vect{r},\vect{r}',\omega)
 \bigr]'_{\omega=\Omega_{nk}}
 \Theta(\tilde{\omega}_{nk}).
\end{align}
Substituting these results into Eq.~(\ref{qf49}), one finds
\begin{equation}
\label{qf52}
\vect{F}_n(\vect{r}_A,\vect{v}_A)
=\vect{F}_n^\mathrm{nr}(\vect{r}_A,\vect{v}_A)
+\vect{F}_n^\mathrm{r}(\vect{r}_A,\vect{v}_A)
\end{equation}
with
\begin{widetext}
\begin{multline}
\label{qf53}
\vect{F}_n^\mathrm{nr}(\vect{r}_A,\vect{v}_A)
=-\frac{\hbar\mu_0}{2\pi}
 \int_0^\infty\dif\xi\,\xi^2
 \bm{\nabla}\trace\bigl\{
 [\bm{\alpha}_n(\vect{v}_A,\mi\xi)
 +\bm{\alpha}_n(\vect{v}_A,-\mi\xi)]
 \sprod\ten{G}^{(1)}(\vect{r}_A,\vect{r}_A,\mi\xi)
 \bigr\}\\
-\frac{\mi\hbar\mu_0}{2\pi}\int_0^\infty\dif\xi\,\xi^2\,
 \bm{\nabla}(\vect{v}_A\sprod\bm{\nabla}')\mathrm{Tr}\bigl\{
 [\bm{\alpha}'_n(\mi\xi)+\bm{\alpha}'_n(-\mi\xi)]
 \sprod\ten{G}^{(1)}(\vect{r}_A,\vect{r}_A,\mi\xi)
 \bigr\}\\
+\frac{\hbar\mu_0}{2\pi}\int_0^\infty\dif\xi\,\xi\,
 (\bm{\nabla}'-\bm{\nabla})\mathrm{Tr}\bigl\{
 [\bm{\alpha}_n(\mi\xi)-\bm{\alpha}_n(-\mi\xi)]
 \sprod\vect{v}_A\vprod\bigl[\bm{\nabla}\vprod
 \ten{G}^{(1)}(\vect{r}_A,\vect{r}_A,\mi\xi)\bigr]
 \bigr\}
\end{multline}
and
\begin{multline}
\label{qf54}
\vect{F}_n^\mathrm{r}(\vect{r}_A,\vect{v}_A)
=\mu_0\sum_k\Theta(\tilde{\omega}_{nk})\Omega_{nk}(\vect{v}_A)
 \bigl[\omega^2\,\bm{\nabla}\tprod\vect{d}_{nk}\sprod
 \ten{G}^{(1)}(\vect{r}_A,\vect{r}_A,\omega)
 \sprod\vect{d}_{kn}\bigr]'_{\omega=\Omega_{nk}}\\
+\mi\mu_0\sum_k\Theta(\tilde{\omega}_{nk})\bigl[\omega^2\,
 \bm{\nabla}(\vect{v}_A\sprod\bm{\nabla}')
 \tprod\vect{d}_{nk}\sprod
 \ten{G}^{(1)}(\vect{r}_A,\vect{r}_A,\omega)
 \sprod\vect{d}_{kn}\bigr]'_{\omega=\Omega_{nk}}\\
+\mi\mu_0\sum_k\Theta(\tilde{\omega}_{nk})\Omega_{nk}
 (\bm{\nabla}'-\bm{\nabla})\tprod\vect{d}_{nk}\sprod
 \vect{v}_A\vprod\bigl[\bm{\nabla}\vprod
 \ten{G}^{(1)}(\vect{r}_A,\vect{r}_A,\Omega_{nk})\bigr]
 \sprod\vect{d}_{kn}
 +\mathrm{C.c.}
\end{multline}
\end{widetext}
Here, 
\begin{equation}
\label{qf61}
\bm{\alpha}_n(\omega)
=\frac{1}{\hbar}\sum_k\biggl[
 \frac{\vect{d}_{kn}\tprod\vect{d}_{nk}}
 {\omega-\Omega_{nk}^\ast}
-\frac{\vect{d}_{nk}\tprod\vect{d}_{kn}}
 {\omega+\Omega_{nk}}\biggr]
\end{equation}
is the polarisability for an atom at rest and
\begin{equation}
\label{qf62}
\bm{\alpha}_n(\vect{v}_A,\omega)
=\frac{1}{\hbar}\sum_k\biggl[
 \frac{\Omega_{nk}^\ast(\vect{v}_A)\vect{d}_{kn}\tprod\vect{d}_{nk}}
 {(\omega-\Omega_{nk}^\ast)^2}
+\frac{\Omega_{nk}(\vect{v}_A)\vect{d}_{nk}\tprod\vect{d}_{kn}}
 {(\omega+\Omega_{nk})^2}\biggr]
\end{equation}
is the correction to this polarisability for a moving atom within
linear order of the atomic velocity.

It is instructive to consider the perturbative limit
$\Omega_{nk}\to\omega_{nk}$ (i.e., $\delta\omega_{n/k}$,
$\Gamma_{n/k}$ $\to 0$). The resonant force can be represented by its
zero-order approximation in $\delta\omega_{n/k}$ and $\Gamma_{n/k}$
which reads
\begin{widetext}
\begin{multline}
\label{qf63}
\vect{F}_n^\mathrm{r}(\vect{r}_A,\vect{v}_A)
=2\mu_0\sum_k\Theta(\omega_{nk})
 [\delta\omega_n(\vect{v}_A)-\delta\omega_k(\vect{v}_A)]
 \bigl[\omega^2\bm{\nabla}\tprod\vect{d}_{nk}\sprod
 \mathrm{Re}\,\ten{G}^{(1)}(\vect{r}_A,\vect{r}_A,\omega)
 \sprod\vect{d}_{kn}\bigr]'_{\omega=\omega_{nk}}\\
-\mu_0\sum_k\Theta(\omega_{nk})
 [\Gamma_n(\vect{v}_A)+\Gamma_k(\vect{v}_A)]
 \bigl[\omega^2\bm{\nabla}\tprod\vect{d}_{nk}\sprod
 \mathrm{Im}\,\ten{G}^{(1)}(\vect{r}_A,\vect{r}_A,\omega)
 \sprod\vect{d}_{kn}\bigr]'_{\omega=\omega_{nk}}\\
-2\mu_0\sum_k\Theta(\omega_{nk})\bigl[\omega^2\,
 \bm{\nabla}(\vect{v}_A\sprod\bm{\nabla}')
 \tprod\vect{d}_{nk}\sprod
 \mathrm{Im}\,\ten{G}^{(1)}(\vect{r}_A,\vect{r}_A,\omega)
 \sprod\vect{d}_{kn}\bigr]'_{\omega=\omega_{nk}}
 \\
-2\mu_0\sum_k\Theta(\omega_{nk})\omega_{nk}
 (\bm{\nabla}'-\bm{\nabla})\tprod\vect{d}_{nk}\sprod
 \vect{v}_A\vprod\bigl[\bm{\nabla}\vprod
 \mathrm{Im}\,\ten{G}^{(1)}(\vect{r}_A,\vect{r}_A,\omega_{nk})\bigr]
 \sprod\vect{d}_{kn}
\end{multline}
The nonresonant velocity-dependent force vanishes to zeroth order in
the frequency shifts and decay rates, in contrast to the force
observed for an atom at rest. The leading nonvanishing contribution
is linear in these quantities and it reads
\begin{multline}
\label{qf64}
\vect{F}_n^\mathrm{nr}(\vect{r}_A,\vect{v}_A)
=-\frac{2\mu_0}{\pi}\sum_k
 \int_0^\infty\dif\xi\,\xi^2\,
 \frac{[\delta\omega_n(\vect{v}_A)-\delta\omega_k(\vect{v}_A)]
 (\omega_{kn}^2-\xi^2)}{(\omega_{kn}^2+\xi^2)^2}\,
 \bm{\nabla}\tprod\vect{d}_{nk}\sprod
 \ten{G}^{(1)}(\vect{r}_A,\vect{r}_A,\mi\xi)
 \sprod\vect{d}_{kn}\\
+\frac{2\mu_0}{\pi}\sum_k\int_0^\infty\dif\xi\,\xi^2\,
 \frac{\omega_{kn}(\Gamma_n+\Gamma_k)(\omega_{kn}^2-3\xi^2)}
 {(\omega_{kn}^2+\xi^2)^3}\,
 \bm{\nabla}(\vect{v}_A\sprod\bm{\nabla}')
 \tprod\vect{d}_{nk}\sprod
 \ten{G}^{(1)}(\vect{r}_A,\vect{r}_A,\mi\xi)
 \sprod\vect{d}_{kn}\\
-\frac{2\mu_0}{\pi}\sum_k\int_0^\infty\dif\xi\,\xi^2\,
 \frac{\omega_{kn}(\Gamma_n+\Gamma_k)}
 {(\omega_{kn}^2+\xi^2)^2}\,
 (\bm{\nabla}'-\bm{\nabla})\tprod\vect{d}_{nk}\sprod
 \vect{v}_A\vprod\bigl[\bm{\nabla}\vprod
 \ten{G}^{(1)}(\vect{r}_A,\vect{r}_A,\mi\xi)\bigr]
 \sprod\vect{d}_{kn}.
\end{multline}
For an isotropic atom, these results reduce to
\begin{multline}
\label{qf65}
\vect{F}_n^\mathrm{r}(\vect{r}_A,\vect{v}_A)
=\frac{2\mu_0}{3}\sum_k\Theta(\omega_{nk})|\vect{d}_{nk}|^2
 [\delta\omega_n(\vect{v}_A)-\delta\omega_k(\vect{v}_A)]
 \bigl[\omega^2\,\bm{\nabla}\operatorname{Tr}
 \mathrm{Re}\,\ten{G}^{(1)}(\vect{r}_A,\vect{r}_A,\omega)
 \bigr]'_{\omega=\omega_{nk}}\\
-\frac{\mu_0}{3}\sum_k\Theta(\omega_{nk})|\vect{d}_{nk}|^2
 [\Gamma_n(\vect{v}_A)+\Gamma_k(\vect{v}_A)]
 \bigl[\omega^2\,\bm{\nabla}\operatorname{Tr}
 \mathrm{Im}\,\ten{G}^{(1)}(\vect{r}_A,\vect{r}_A,\omega)
 \bigr]'_{\omega=\omega_{nk}}\\
-\frac{2\mu_0}{3}\sum_k\Theta(\omega_{nk})|\vect{d}_{nk}|^2
 \bigl[\omega^2\bm{\nabla}(\vect{v}_A\sprod\bm{\nabla}')
 \operatorname{Tr} 
 \mathrm{Im}\,\ten{G}^{(1)}(\vect{r}_A,\vect{r}_A,\omega)
 \bigr]'_{\omega=\omega_{nk}}\\
-\frac{2\mu_0}{3}\sum_k\Theta(\omega_{nk})\omega_{nk}|\vect{d}_{nk}|^2
 (\bm{\nabla}'-\bm{\nabla})\operatorname{Tr}\bigl\{
 \vect{v}_A\vprod\bigl[\bm{\nabla}\vprod
 \mathrm{Im}\,\ten{G}^{(1)}(\vect{r}_A,\vect{r}_A,\omega_{nk})\bigr]
 \bigr\}
\end{multline}
and
\begin{multline}
\label{qf66}
\vect{F}_n^\mathrm{nr}(\vect{r}_A,\vect{v}_A)
=-\frac{2\mu_0}{3\pi}\sum_k|\vect{d}_{nk}|^2
 \int_0^\infty\dif\xi\,\xi^2\,
 \frac{[\delta\omega_n(\vect{v}_A)-\delta\omega_k(\vect{v}_A)]
 (\omega_{kn}^2-\xi^2)}{(\omega_{kn}^2+\xi^2)^2}\,
 \bm{\nabla}\operatorname{Tr}
 \ten{G}^{(1)}(\vect{r}_A,\vect{r}_A,\mi\xi)\\
+\frac{2\mu_0}{3\pi}\sum_k|\vect{d}_{nk}|^2
 \int_0^\infty\dif\xi\,\xi^2\,
 \frac{\omega_{kn}(\Gamma_n+\Gamma_k)(\omega_{kn}^2-3\xi^2)}
 {(\omega_{kn}^2+\xi^2)^3}\,
 \bm{\nabla}(\vect{v}_A\sprod\bm{\nabla}')
 \operatorname{Tr}\ten{G}^{(1)}(\vect{r}_A,\vect{r}_A,\mi\xi)\\
-\frac{2\mu_0}{3\pi}\sum_k|\vect{d}_{nk}|^2
 \int_0^\infty\dif\xi\,\xi^2\,
 \frac{\omega_{kn}(\Gamma_n+\Gamma_k)}
 {(\omega_{kn}^2+\xi^2)^2}\,
 (\bm{\nabla}'-\bm{\nabla})\operatorname{Tr}\bigl\{
 \vect{v}_A\vprod\bigl[\bm{\nabla}\vprod
 \ten{G}^{(1)}(\vect{r}_A,\vect{r}_A,\mi\xi)\bigr]
 \bigr\}.
\end{multline}
\end{widetext}


\subsection{Motion parallel to a planar interface}
\label{Sec3.3}

\begin{figure}[!t!]
\includegraphics[width=0.8\linewidth]{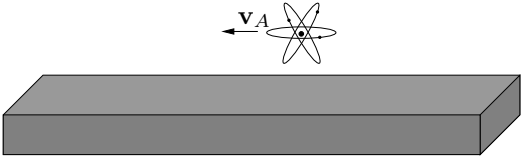}
\caption{\label{fig:planar} Atom moving with a velocity $\vect{v}$
near a planar surface.}
\end{figure}

Up until this point, all results are valid for arbitrary geometries.
In order to gain physical insight, we restrict ourselves to the
generic quantum friction scenario of an atom moving parallel
[$\vect{v}_A=\vect{v}_\parallel=(v_x,v_y,0)^\trans$] to a
homogeneous dielectric or metal of permittivity $\varepsilon(\omega)$
whose plane surface defines the $(x,y)$-plane (see
Fig.~\ref{fig:planar}). The Weyl expansion of the Green tensor
\begin{equation}
\label{qf67}
\ten{G}(\vect{r},\vect{r}',\omega) = \int \frac{\dif^2k_\|}{(2\pi)^2}
\me^{\mi\mathbf{k}_\|\cdot(\bm{\rho}-\bm{\rho}')}
\bm{G}(\vect{k}_\|,z,z',\omega) 
\end{equation}
with $\mathbf{r}=(\bm{\rho},z)$ can then be used to calculate
explicit expressions for the terms that contribute to the velocity
dependent force. The relevant Weyl components
$\bm{G}(\mathbf{k}_\|,z,z',\omega)$ of the Green tensor for $z,z'>0$
are given by [$G_{ij}\equiv G_{ij}(\mathbf{k}_\|,z,z',\omega)$]
%
\begin{eqnarray}
\label{qf68}
G_{xx} &\!=&\! \frac{\mi}{2k_{z}}\,
\me^{\mi k_{z}(z+z')} \biggl[ r_s\, \frac{k_y^2}{k_\|^2}
-r_p\, \frac{k_{z}^2k_x^2}{k^2k_\|^2} \biggr] ,\\
\label{qf69}
G_{xy} &\!=&\! \frac{\mi}{2k_{z}}\,
\me^{\mi k_{z}(z+z')} \biggl[ -r_s\, \frac{k_xk_y}{k_\|^2}
-r_p\, \frac{k_{z}^2k_xk_y}{k^2k_\|^2}  \biggr] ,\quad \\
\label{qf70}
G_{xz} &\!=&\! -\frac{\mi}{2k_{z}}\,
\me^{\mi k_{z}(z+z')} r_p\, \frac{k_{z}k_x}{k^2},\\
\label{qf71}
G_{zz} &\!=&\! \frac{\mi}{2k_{z}}\,
\me^{\mi k_{z}(z+z')} r_p\, \frac{k_\|^2}{k^2}\,,
\end{eqnarray}
with
\begin{equation}
\label{qf71b}
r_s =
\frac{k_z-k_{1z}}{k_z+k_{1z}}\,,
 \quad
r_p =
\frac{\varepsilon(\omega)k_z-k_{1z}}{\varepsilon(\omega)k_z+k_{1z}}
\end{equation}
being the Fresnel reflection coefficients of the surface for $s$- and
$p$-polarised waves [$k^2=\omega^2/c^2$,
$k_1^2=\varepsilon(\omega)\omega^2/c^2$,
$k_{(1)z}^2=k_{(1)}^2-k_\|^2$]. The other components of the Green
tensor can be obtained by using the reciprocity
condition $\bm{G}(\mathbf{r},\mathbf{r}',\omega)$
$=\bm{G}^T(\mathbf{r}',\mathbf{r},\omega)$, which translates into
$\bm{G}(\mathbf{k}_\|,z,z',\omega)$
$=\bm{G}^T(-\mathbf{k}_\|,z',z,\omega)$, and the replacement rules
$G_{yy}=G_{xx}(k_x\leftrightarrow k_y)$,
$G_{yz}=G_{xz}(k_x\leftrightarrow k_y)$. 

For the assumed motion parallel to the surface, the velocity-dependent
shifts and rates vanish $\delta\omega_n^k(\vect{r}_A,\vect{v}_A)=
\Gamma_n^k(\vect{r}_A,\vect{v}_A)=0$ (cf.~the remark at the end of
Sec.~\ref{Sec3.1}), and so do the generalised Doppler contributions
to the resonant force~(\ref{qf65}) (first two terms) and the
non-resonant force~(\ref{qf66}) (first term). To calculate the
delay and R\"{o}ntgen contributions, we require second derivatives of
the Green tensor as given above. It is useful to note that all those
derivatives vanish that do not contain an even number for each of the
cartesian indices $(x,y,z)$. For example, terms such as
$\partial_x\partial_yG_{xx}$ or $\partial_y\partial_yG_{xz}$ will not
contribute whereas terms such as $\partial_x\partial_yG_{xy}$ or
$\partial_y\partial_yG_{zz}$ will. For simplicity, we restrict our
attention to the nonretarded or near-field limit, where the dominant
contribution to the Green tensor is due to evanescent waves with
$k_{1z}\simeq k_z\simeq\mi k_\parallel$. With this replacement, we
have
\begin{equation}
\label{qf72b}
r_s=0\,,
 \quad
r_p =
\frac{\varepsilon(\omega)-1}{\varepsilon(\omega)+1}
\end{equation}
and Eqs.~(\ref{qf67})--(\ref{qf71}) lead to
\begin{gather}
\label{qf73b}
\vect{\nabla}(\vect{v}_\parallel\sprod\vect{\nabla}')
 \operatorname{Tr}\ten{G}^{(1)}(\vect{r}_A,\vect{r}_A,\omega)
=\frac{3c^2\vect{v}_\parallel}{16\pi\omega^2z_A^5}\,
 \frac{\varepsilon(\omega)-1}{\varepsilon(\omega)+1}\,,\\
\label{qf72}
(\vect{\nabla}'-\vect{\nabla})\operatorname{Tr}\bigl\{
 \vect{v}_\parallel\vprod\bigl[\vect{\nabla}\vprod
 \ten{G}^{(1)}(\vect{r}_A,\vect{r}_A,\omega)\bigr]\bigr\}
=\bm{0}. 
\end{gather}
In the near-field limit, the R\"{o}ntgen contribution hence also
vanishes and quantum friction is entirely due to the delay
effect. 

Substituting Eqs.~(\ref{qf73}) and (\ref{qf72}) into
Eqs.~(\ref{qf65}) and (\ref{qf66}), we find the friction force
\begin{multline}
\label{qf73}
\vect{F}_n^\mathrm{r}(\vect{r}_A,\vect{v}_\parallel)
 =-\frac{\vect{v}_\parallel}{4\pi\varepsilon_0z_A^5}
 \sum_k\Theta(\omega_{nk})|\vect{d}_{nk}|^2\\
\times\biggl[\frac{\mathrm{Im}\varepsilon(\omega)}
 {|\varepsilon(\omega)+1|^2}\biggr]'_{\omega=\omega_{nk}}
\end{multline}
and
\begin{multline}
\label{qf74}
\vect{F}_n^\mathrm{nr}(\vect{r}_A,\vect{v}_\parallel)
= -\frac{\vect{v}_\parallel}{8\pi\varepsilon_0z_A^5}
 \sum_k|\vect{d}_{nk}|^2\omega_{kn}(\Gamma_n+\Gamma_k)\\
\times\int_0^\infty\dif\xi\,
 \frac{\omega_{kn}^2-3\xi^2}{(\omega_{kn}^2+\xi^2)^3}\,
 \frac{\varepsilon(\mi\xi)-1}{\varepsilon(\mi\xi)+1}\,.
\end{multline}
If we further assume a single-resonance Drude--Lorentz model for the
permittivity,
\begin{equation}
\label{qf75}
\varepsilon(\omega) = 1+\frac{\omega_\mathrm{P}^2}
{\omega_\mathrm{T}^2-\omega^2-2\mi\gamma\omega}\,, 
\end{equation}
with Plasma frequency $\omega_\mathrm{P}$, transverse resonance
frequency $\omega_\mathrm{T}$ and line width $\gamma$, we find that
for a weakly absorbing medium ($\gamma\ll\omega_\mathrm{P,T}$) the
resonant and nonresonant forces are given by
\begin{multline}
\label{qf76}
\vect{F}_n^\mathrm{r}(\vect{r}_A,\vect{v}_\parallel)
 =\frac{\vect{v}_\parallel}{8\pi\varepsilon_0z_A^5}
 \sum_k\Theta(\omega_{nk})|\vect{d}_{nk}|^2\\
\times\frac{\gamma\omega_\mathrm{P}^2
 (\omega_\mathrm{S}^2+3\omega_{nk}^2)}
 {(\omega_{nk}^2-\omega_\mathrm{S}^2)^3}
\end{multline}
and
\begin{multline}
\label{qf77}
\vect{F}_n^\mathrm{nr}(\vect{r}_A,\vect{v}_\parallel)
= -\frac{\vect{v}_\parallel}{32\pi\varepsilon_0z_A^5}
 \sum_k|\vect{d}_{nk}|^2\\
\times\frac{\mathrm{sign}(\omega_{kn})
 (\Gamma_n+\Gamma_k)\omega_\mathrm{P}^2}
 {\omega_\mathrm{S}(|\omega_{kn}|+\omega_\mathrm{S})^3}\,.
\end{multline}
($\omega_\mathrm{S}=\sqrt{\omega_\mathrm{T}^2+\omega_\mathrm{P}^2/2}$,
surface plasmon frequency).

Let us discuss our results. We first note that in a quantum friction
scenario of an atom moving parallel to a plane surface, a generalised
Doppler effect does not contribute to the velocity-dependent force;
this will be different for an atom moving perpendicularly towards the
surface. In the near-field limit, the magnetic R\"{o}ntgen coupling
becomes becomes negligible as well; it will become relevant for larger
distances. Near-field quantum friction forces are hence dominantly
caused by a delay effect.

For a ground-state atom, only a nonresonant force component~(\ref{qf77})
is present. With both $\omega_{k0}$ and $\Gamma_k$ being positive
quantities, $\vect{F}_n^\mathrm{nr}(\vect{r}_A,\vect{v}_\parallel)$ is
strictly antiparallel to the velocity and hence presents a genuine
friction force. Note that this force is proportional to the rates of
spontaneous decay $\Gamma_k$, the absorption parameters of the atom.
In the near-field limit, these decay rates are given by
\cite{Yeung96,Henkel99,Heating}
\begin{multline}
\label{qf78}
\Gamma_n=\sum_k\Gamma_{nk}
=\sum_k\Theta(\omega_{nk})\frac{|\vect{d}_{nk}|^2}
 {6\pi\hbar\varepsilon_0z_A^3}\,
 \frac{\mathrm{Im}\varepsilon(\omega_{nk})}
 {|\varepsilon(\omega_{nk})+1|^2}\\
=\sum_k\Theta(\omega_{nk})\frac{|\vect{d}_{nk}|^2}
 {12\pi\hbar\varepsilon_0z_A^3}\,
 \frac{\gamma\omega_{nk}\omega_\mathrm{P}^2}
 {(\omega_{nk}^2-\omega_\mathrm{S}^2)^2}\,,
\end{multline}
Inserting this into Eq.~(\ref{qf77}) yields a friction force that is
extremely short-ranged and falls off as $z_A^{-8}$. This is in
contrast to previous theories \cite{Dedkov} that predict a
$z_A^{-5}$-scaling, resulting from a disregard of the
distance-dependence of the spontaneous decay rate.

For an excited atom, resonant forces arise as a consequence of
possible transitions to lower lying atomic energy levels. They will
dominate the velocity-dependent force, in particular if one of them
is near-resonant with the surface plasmon frequency $\omega_S$.
Depending on whether the respective atomic transition frequency
$\omega_{nk}$ is smaller or greater than the surface plasmon
frequency, the velocity-dependent resonant force will either be a
decelerating friction force antiparallel to the velocity, or it may be
an quantum acceleration force parallel to the velocity. This can be
qualitatively understood from an energy consideration. The energy
$\hbar\omega_{nk}$ emitted during a downward transition of the atom is
resonantly absorbed by the surface and leads to the excitation of a
surface plasmon with energy $\hbar\omega_\mathrm{S}$. The energy
difference between these two reservoirs leads to a change in the
atom's kinetic energy. If the emitted energy is smaller than the
absorbed one, the atom has to decelerate; if the emitted energy is
greater than the absorbed one, the atom will accelerate. Note also
that the off-resonant contribution~(\ref{qf77}) consists of strictly
accelerating downward contributions as well as strictly decelerating
upward contributions which, as before, have a $z_A^{-8}$-scaling and
can be safely neglected with regard to the $z_A^{-5}$-scaling of the
resonant forces. It is known from previous studies \cite{Boussiakou}
that the rate of spontaneous decay increases for atoms in motion. This
mechanism leads to a more rapid dissipation of the internal energy
initially stored in the atom, restricting the life time of resonant
forces; it needs to be taken into account in a more quantitative
analysis of energy conservation. 


\section{Examples}
\label{Sec4}

In order to illustrate the effect of velocity-dependent forces on
atoms, we present a selection of numerical examples. We will
concentrate on forces that are dominated by a single atomic transition
between a ground state $|0\rangle$ and an excited state $|1\rangle$
with frequency $\omega_A$ and (isotropic) dipole matrix element $d$.
In this case, the nonresonant ground-state force, Eq.~(\ref{qf77}),
reduces to the simpler expression
\begin{equation}
\vect{F}_0(\vect{r}_A,\vect{v}_\parallel)
= -\frac{\vect{v}_\parallel d^2}{32\pi\varepsilon_0z_A^5}\,
\frac{\Gamma\omega_\mathrm{P}^2}
{\omega_\mathrm{S}(\omega_A+\omega_\mathrm{S})^3} \,,
\end{equation}
where the decay rate (\ref{qf78}) now reads
\begin{equation}
\label{eq:gammatwolevel}
\Gamma=\frac{d^2}{12\pi\hbar\varepsilon_0z_A^3}\,
\frac{\gamma\omega_A\omega_\mathrm{P}^2}
{(\omega_A^2-\omega_\mathrm{S}^2)^2} \,.
\end{equation}
The excited-state force is dominated by the resonant force component,
$\vect{F}_1(\vect{r}_A,\vect{v}_\parallel)$
$=\vect{F}_1^\mathrm{nr}(\vect{r}_A,\vect{v}_\parallel)$
$+\vect{F}_1^\mathrm{r}(\vect{r}_A,\vect{v}_\parallel)$
$\approx\vect{F}_1^\mathrm{r}(\vect{r}_A,\vect{v}_\parallel)$,
\begin{equation}
\label{eq:F1twolevel}
\vect{F}_1(\vect{r}_A,\vect{v}_\parallel)
 =\frac{\vect{v}_\parallel d^2}{8\pi\varepsilon_0z_A^5}\,
\frac{\gamma\omega_\mathrm{P}^2(\omega_\mathrm{S}^2+3\omega_A^2)}
{(\omega_A^2-\omega_\mathrm{S}^2)^3} \,.
\end{equation}

As a first example, we consider a ground-state ${}^{87}$Rb atom moving
parallel to a gold surface. We consider the lowest electronic
transition
$D_2(5^2\mathrm{S}_{1/2}\!\rightarrow\!5^2\mathrm{P}_{3/2})$ with
wavelength $\lambda_A=780\,\mbox{nm}$ ($\omega_A\!=\!2.41\times
10^{15}\,\mbox{rad s}^{-1}$) \cite{rubidium} and dipole moment
$d\!=\!4.23ea_0\!=\!3.58\times 10^{-29}\,\mbox{Cm}$ \cite{Steck}. The
permittivity of gold may be characterised by a plasma frequency
$\omega_\mathrm{P}\!=\!1.37\times 10^{16}\,\mbox{rad s}^{-1}$ and an
absorption parameter $\gamma\!=\!4.12\times 10^{13}\,\mbox{rad
s}^{-1}$ \cite{Heating}. Note that the transverse resonance frequency
vanishes for metals, $\omega_\mathrm{T}\!=\!0$, so that the surface
plasmon resonance is located at
$\omega_\mathrm{S}\!=\!\omega_\mathrm{P}/\sqrt{2}$. With these
parameters, we find a deceleration of the rubidium atom as
($m_{^{87}\mathrm{Rb}}=1.44\times 10^{-25}\,\mbox{kg}$)
\begin{equation}
\vect{a}_\|= -\vect{v}_\|\bigl( 9.6\,\mbox{s}^{-1}\bigr)
\left[\frac{1\,\mbox{nm}}{z_A}\right]^8 \,.
\end{equation}
The force is extremely short-ranged, and is negligible for any
reasonable values of the velocity and atom-surface distance.

In contrast, for an excited rubidium atom with the same data as above,
the deceleration becomes
\begin{equation}
\label{eq:metalfriction}
\vect{a}_\| = -\vect{v}_\| \bigl( 5.0\times 10^4\,\mbox{s}^{-1} \bigr)
\left[\frac{1\,\mbox{nm}}{z_A}\right]^5 \,.
\end{equation}
In comparison to the ground-state force, excited-state quantum
friction is strongly enhanced and has a much longer range. For an
atomic velocity of $v\!=\!200\,\mbox{ms}^{-1}$, the deceleration at an
atom-surface distance $z_A\!=\!10\,\mbox{nm}$ can be as large as
$a\!=\!-100\,\mbox{ms}^{-2}$. Even at $z_A\!=\!100\,\mbox{nm}$ the
deceleration is still $a\!=\!-10^{-3}\,\mbox{ms}^{-2}$.

Results for other atoms and metallic surfaces can be easily obtained
by noting that in most cases, the relevant atomic transition frequency
is much smaller than the surface plasmon frequency of the metal, hence
$\omega_A\!\ll\!\omega_\mathrm{S}$. Under this approximation, the
excited-state force (\ref{eq:F1twolevel}) and the decay rate
(\ref{eq:gammatwolevel}) read
\begin{equation}
\vect{F}_1(\vect{r}_A,\vect{v}_\|) \approx
-\frac{\vect{v}_\|d^2}{2\pi\varepsilon_0z_A^5}\,
\frac{\gamma}{\omega_\mathrm{P}^2}
\end{equation}
and
\begin{equation}
\Gamma \approx \frac{d^2\omega_A}{3\pi\hbar\varepsilon_0z_A^3}\,
\frac{\gamma}{\omega_\mathrm{P}^2}\,.
\end{equation}
Typical values for the material parameter $\omega_\mathrm{P}^2/\gamma$
are tabulated in Ref.~\cite{Heating}. Note that unless the excitation
is maintained by continuous repumping, the excited-state force only
acts during a time interval $\Delta t\approx\Gamma^{-1}$. The relative
velocity change during this time is approximately
\begin{equation}
\frac{\Delta v}{v} \approx \frac{F_1}{\Gamma mv} \approx
-\frac{3\hbar}{2m_A\omega_Az_A^2}\,.
\end{equation}
In this limit ($\omega_A\!\ll\!\omega_\mathrm{S}$), the relative
change in velocity is independent of the strength of the atomic dipole
transition and all material parameters.

Upon inspection of the excited-state force (\ref{eq:F1twolevel}) one
notices that this force can be resonantly enhanced if an atomic
transition matches the frequency of a surface plasmon resonance. An
example of such a close match has been pointed out in
Refs.~\cite{cesiumdata,Gorza06} and involves a sapphire substrate with
its principal surface plasmon at
$\lambda_\mathrm{S}\!=\!12.21\,\mu\mbox{m}$
($\omega_\mathrm{S}\!=\!1.54\times10^{14}\,\mbox{rad s}^{-1}$) and
the $6D_{3/2}\!\rightarrow\!7P_{1/2}$ transition in ${}^{133}$Cs with
a wavelength of $\lambda_A=12.15\,\mu\mbox{m}$
($\omega_A\!=\!1.55\times10^{14}\,\mbox{rad s}^{-1}$). Near this
plasmon resonance, the permittivity of sapphire is well approximated
by \cite{Gorza06}
\begin{equation}
\label{sapp}
\varepsilon_\mathrm{sapp}(\omega) = \eta +
\frac{\eta\omega_\mathrm{P}^2}
{\omega_\mathrm{T}^2-\omega^2-2\mi\gamma\omega } 
\end{equation}
with $\eta=2.71$, $\omega_\mathrm{P}\!=\!0.84\omega_\mathrm{S}%
\!=\!1.29\times 10^{14}\,\mbox{rad s}^{-1}$,
$\omega_\mathrm{T}=0.70\omega_\mathrm{S}%
\!=\!1.08\times 10^{14}\,\mbox{rad s}^{-1}$, and 
$\gamma=7.5\times 10^{-3}\omega_\mathrm{S}%
\!=\!1.16\times 10^{12}\,\mbox{rad s}^{-1}$; note that
$\omega_S=\sqrt{\omega_\mathrm{T}^2+\omega_\mathrm{P}\eta/(\eta+1)}$.
With this model, and introducing the atom-plasmon detuning
$\delta=\omega_A-\omega_\mathrm{S}$, we find that the
force~(\ref{qf73}) and the decay rate~(\ref{qf78}) in the vicinity of
the surface-plasmon resonance may be given as
($|\delta|,\gamma\ll\omega_\mathrm{S}$)
\begin{equation}
\vect{F}_1(\vect{r}_A,\vect{v}_\|) \approx
\frac{\vect{v}_\|d^2}{4\pi\varepsilon_0z_A^5}\,
\frac{\eta}{(\eta+1)^2}\,\frac{\omega_\mathrm{P}^2}{\omega_\mathrm{S}}
\,\frac{\gamma\delta}{(\delta^2+\gamma^2)^2} 
\end{equation}
and
\begin{equation}
\Gamma \approx\frac{d^2}{12\pi\hbar\varepsilon_0z_A^3}\,
\frac{\eta}{(\eta+1)^2}\,\frac{\omega_\mathrm{P}^2}{\omega_\mathrm{S}}
\,\frac{\gamma}{\delta^2+\gamma^2}\,.
\end{equation}

With the dipole moment of the abovementioned transition being
$d\!=\!5.85\times 10^{-29}\,\mbox{Cm}$ \cite{Lindgard}, one finds
($m_{^{133}\mathrm{Cs}}=2.21\times 10^{-25}\,\mbox{kg}$)
\begin{equation}
\vect{a}_\| = +\vect{v}_\|\bigl(7.1\times10^{11}\mbox{s}^{-1}\bigr)
\left[\frac{1\,\mbox{nm}}{z_A}\right]^5\,.
\end{equation}
Compared with the result (\ref{eq:metalfriction}) for the
excited-state force near a metal, we find a significantly enhanced
force. Note also that, because $\omega_A>\omega_S$ (i.e., $\delta>0$),
the force is accelerating rather than decelerating. As a numerical
example, for a particle velocity of $v=100\,\mbox{ms}^{-1}$ and an
atom-surface distance of $z_A=100\,\mbox{nm}$, one would observe an
acceleration of $a=7\times10^4\,\mbox{ms}^{-2}$. As before, without
continuous repumping this force acts only for a very short time,
leading to a net relative change in velocity
\begin{equation}
\frac{\Delta v}{v} \approx \frac{3\hbar}{mz_A^2}\,
\frac{\delta}{\delta^2+\gamma^2}\,.
\end{equation}


\vspace*{4ex}

\section{Summary}
\label{Sec:conclusions}

We have shown that atoms or molecules in relative motion with respect
to a dielectric surface experience velocity-dependent CP forces.
Solving the coupled atom-field dynamics for a slowly moving atom, we
have found an expression for the linearised velocity-dependent force
on an atom in an arbitrary incoherent internal quantum state moving
near an arbitrary arrangement of magnetoelectric bodies. In general,
three effects contribute to the velocity-dependent Casimir--Polder
force: a generalised Doppler effect due to the velocity-dependence of
the atomic transition frequencies, the delay between the emission and
reabsorption of photons by the atom and the R\"ontgen interaction,
i.e., to the coupling of the current density associated with the
atomic motion to the magnetic field.

In order to illustrate the general theory, we have studied the
near-field force on an atom that moves parallel to a planar dielectric
or metallic surface. Due to the translational invariance of the
system, the Doppler term does not contribute in this case.
Furthermore, the delay term dominates over the R\"ontgen term. For a
ground-state atom the force is a genuine friction force, i.e. a force
antiparallel to its velocity. It is proportional to the atomic
linewidth and hence very small. In contrast, excited-state atoms can
be either decelerated or accelerated depending on the relative
magnitude of their transition frequency with respect to the
characteristic frequency of the substrate material. For metals, the
force is always decelerating while for dielectric substrates with
sufficiently small surface plasmon frequency, acceleration of
excited-state atoms can be realised.

In addition, the force on such atoms is strongly enhanced when atom
and substrate are near-resonant. Much stronger enhancement can be
achieved when the atom moves through resonator structures, in close
analogy to the stationary case \cite{Ellingsen09}.


\acknowledgments
The authors gratefully acknowledge discussions with T.~Freegarde and
P.~Horak. This work was supported by the UK Engineering and Physical
Sciences Research Council and the Alexander~von~Humboldt foundation.


\end{document}